# XML Entity Architecture for Efficient Software Integration

Amol Patwardhan[#1], Rahul Patwardhan[*2]

[#] *Department of Development and Operations, Asset Mark Inc*

[1]`amolpatty@gmail.com`

[2]`rahul.patwardhan.2006@gmail.com`

*Abstract*— This paper proposed xml entities based architectural implementation to improve integration between multiple third party vendor software systems with incompatible xml schema. The xml entity architecture implementation showed that the lines of code change required for mapping the schema between in house software and three other vendor schema, decreased by 5.2%, indicating an improvement in quality. The schema mapping development time decreased by 3.8% and overall release time decreased by 5.3%, indicating an improvement in productivity. The proposed technique proved that using xml entities and XSLT transforms is more efficient in terms of coding effort and deployment complexity when compared to mapping the schema using object oriented scripting language such as C#.

## I. Introduction

Business logic layer consists of classes representing business entities. The business entities map to a one or more tables and view. In the model view controller architecture, the model is the equivalent of business entities. Sometimes there is a business need to integrate the software with another 3rd party vendor system. Such an integration requires mapping the vendor schema with the business entities of the host and in house software system. A limitation of class level representation for business entities is that the mapping process involves additional processing overhead. This paper proposes the xml business entities based architecture to solve the problem of integration overhead with 3rd party vendor software systems. We used 3 internal projects to compare baseline and new method implementation data. Each project had been implemented in asp.net, C# using N-tier architecture. The business entities were represented by C# classes. The mapping code was implemented using C#.

## II. Related work

The Microsoft Application Architecture Guide [1] described the scenarios when xml business entities can be used. It recommended using xml entities when the presentation layer requires it or when there is specific logic built around the xml element. The guide also recommends using custom xml objects when the system to system message exchange is being done using xml. It also warned against possible high memory usage while processing huge amount of in memory-xml data. We hypothesize that the xml based business entities can be used more extensively for a broader range of scenarios and even as a replacement for custom objects defined by classes. We believe that the implementation advantages exceed the performance concerns.

## III. Methodology

To measure the baseline data, we used following metrics:
a) Number of project files modified.
b) Number of defects fixed per release.
c) Development time to implement the mapping.
d) Development time to implement features and changes to schema.
e) Release time for fixing defects.
f) Deployed lines of code.
g) Memory footprint of data.
h) Deserialization and Serialization time.

First, we measured the metric for an existing project in the solution. The internal project code was DT. The project had been implemented using ASP.NET C#, Web forms and SQL 2008. We obtained the metrics from existing TFS change-sets, shelves, check-ins, review emails, project release documents, project plan and quality assurance tracking system (QA). After the baseline data was established we used the proposed xml entities based architecture for the next implementation of vendor integration modules that involved mapping third party xml schema with the in-house xml schema. The same set of metrics were tracked to obtain a comparison with the baseline. The empirical data was obtained over a period of 14 months and included 4 releases and 8 maintenance release cycles. The data was aggregated for 2 new projects (internal project code R1 and CUDL) that implemented the proposed architecture and maintenance cycles for the existing project (internal code DT) that used legacy mapping code.

IV. RESULT

The number of files modified per release decreased when the proposed xml based entities architecture was used in the project as compared to the implementation that used class based business entities representation, as shown in Fig. 1.

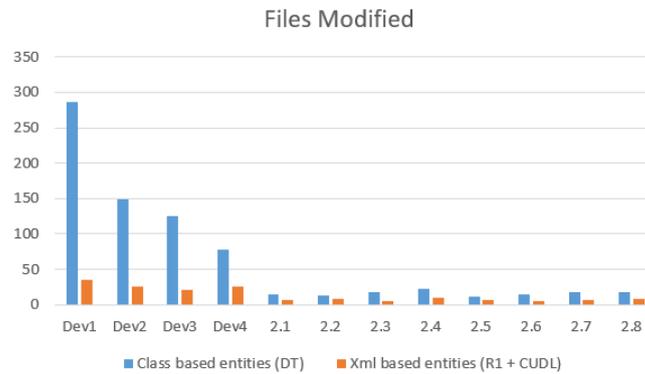

Fig. 1 Number of files modified per release

Fig. 2 shows that the number of defects fixed per release increased when the xml based entities architecture was used. This indicates an improvement in productivity per available development resource (programmers, quality assurance staff).

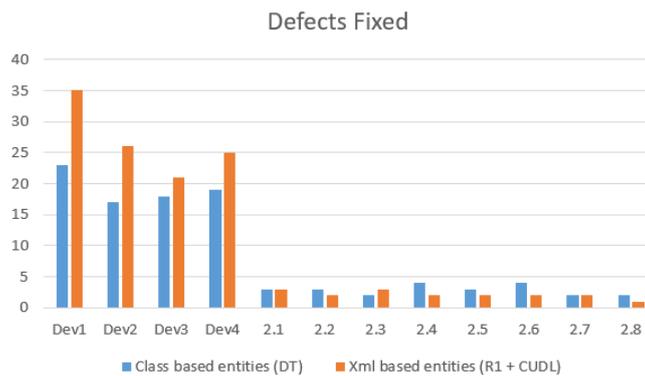

Fig. 2 Number of defects fixed per release

Fig. 3 shows that the schema mapping duration per release decreased when the xml based entities architecture was used. This indicates that the implementation effort reduced and contributed in overall productivity of the programmer.

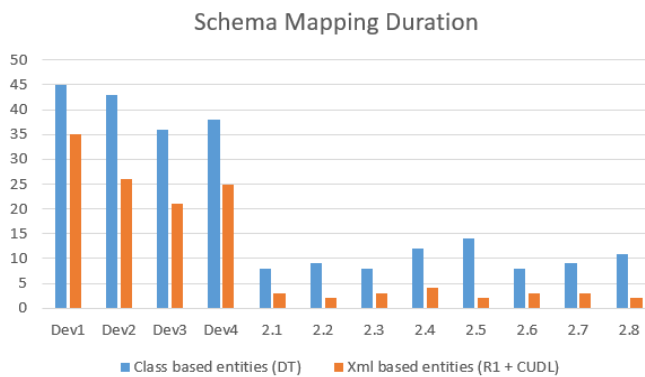

Fig. 3 Schema mapping duration in person-days per release

Fig. 4 shows that the code change effort in person-days per release decreased when the xml based entities architecture was used. This indicates an improvement in productivity per available development resource (programmers, quality assurance staff).

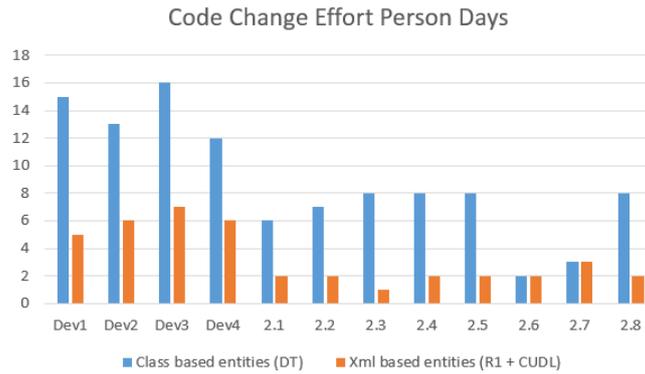

Fig. 4 Code change effort in person-days per release

Fig. 5 shows that the maintenance release duration decreased when the xml based entities architecture was used. This indicates an improvement in efficiency of release management process when the proposed architecture was used. The decrease in release duration between process that used legacy architecture and the new xml based entities architecture was 5.3%.

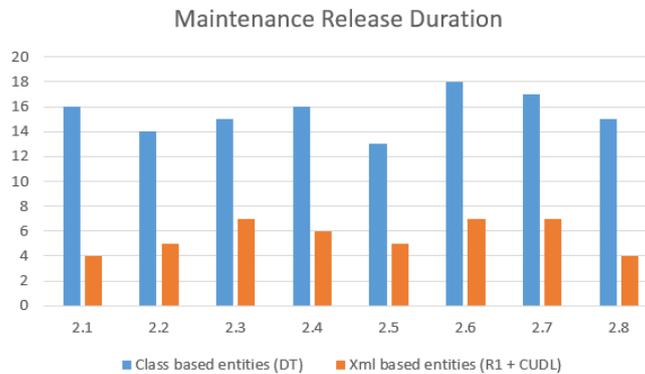

Fig. 5 Maintenance release duration in person-days per release

Fig. 6 shows that the number of deployed lines of code per release increased when the xml based entities architecture was used. This indicates a reduction in code churn and thus decrease in the amount of defect regression and an improvement in the quality of the code. The xml based entities architecture also required very few lines of code changes and was highly flexible in terms of build and deploy. It also reduced the lines of code that had to be code reviewed.

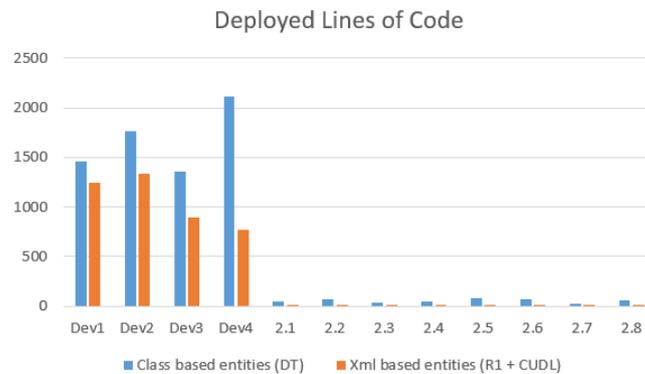

Fig. 6 Deployed lines of code per release

Fig. 7 shows that the available memory in megabytes (MB) per release increased when the xml based entities architecture was used. This indicates that the memory required by the xml entities based architecture was higher than the class based business entities representation. But this increase was only 0.48% and could be easily handled by available memory resources on the production as well as development servers.

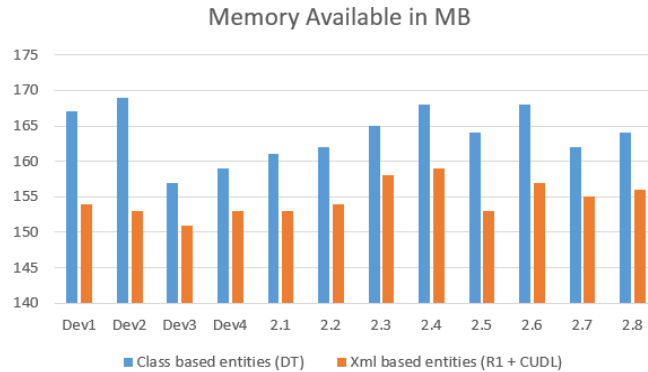

Fig. 7 Memory available in MB per release

Fig. 8 shows that the processing time for mapping the schema per release decreased when the xml based entities architecture was used. This indicates an improvement in processing performance and faster software response. The improvement was 0.875% and was significant in terms of business-to-business (B2B) system communication over the network.

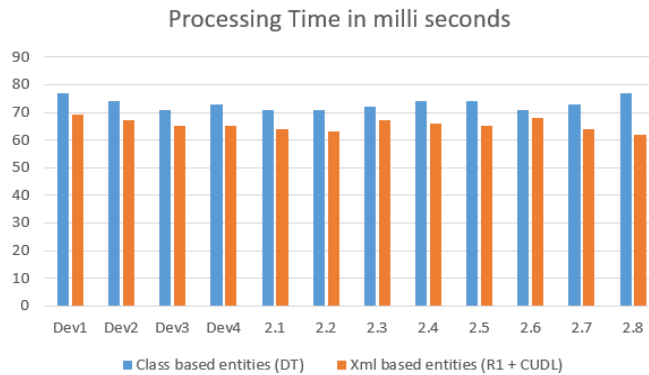

Fig. 8 Processing time in milli-seconds per release

## V. Conclusions

Enterprise level software development requires integration with third party APIs, services and software systems over the internet. The proposed xml entities based software implementation improves the integration process in terms of coding effort and release time when compared to integration with software containing classes to represent business entities. The change management and bug fix process also greatly improves in terms of coding and testing effort when xml based entities are mapped to third party vendor software schema using XSLT transforms. The xml based entities architecture also provides a foundation to extend the code with minimal change. As a future scope the feasibility of the xml entities architecture in document generation and code comprehension must be investigated.